\documentclass[manuscript]{raa}            

\usepackage{graphicx,times,natbib,longtable,rotating}             
\input{epsf.sty}                        
\input{psfig.sty}                       

\begin{document}

   \title{The THU-NAOC Transient Survey: the Performance and the First-year Result
}

   \volnopage{Vol.0 (200x) No.0, 000--000}      
   \setcounter{page}{1}          

   \author{Tianmeng Zhang,
      \inst{1}
      Xiaofeng Wang,
      \inst{2}
      Junchen Chen,
      \inst{2}
      Jujia Zhang,
      \inst{3,4,5}
      Li Zhou,
      \inst{2}
      Wenxiong Li,
      \inst{2}
      Qing Liu,
      \inst{2}
      Jun Mo,
      \inst{2}
      Kaicheng Zhang,
      \inst{2}
      Xinyu Yao,
      \inst{2,6}
      Xulin Zhao,
      \inst{2}
      Xu Zhou,
      \inst{1}
      Jundan Nie,
      \inst{1}
      Fang Huang,
      \inst{2,7}
      Zhaoji Jiang,
      \inst{1}
      Jun Ma,
      \inst{1}
      Lingzhi Wang,
      \inst{1}
      Chao Wu,
      \inst{8}
      Zhimin Zhou,
      \inst{1}
      Hu Zou,
      \inst{1}
      Lifan Wang
      \inst{6}
         }

   \institute{Key Laboratory of Optical Astronomy,
National Astronomical Observatories, Chinese Academy of Sciences, Beijing 100012, China; zhangtm@nao.cas.cn\\
        \and
        Physics Department and Tsinghua Center for Astrophysics (THCA), Tsinghua University, Beijing 100084, China; wang\_xf@mail.tsinghua.edu.cn
        \and
        Yunnan Observatories (YNAO), Chinese Academy of Sciences, Kunming 650011, China
        \and
        Key Laboratory for the Structure and Evolution of Celestial Objects, Chinese Academy of Sciences, Kunming 650011, China
        \and
        University of Chinese Academy of Sciences, Beijing 100049, China
        \and
        Purple Mountain Observatory, Chinese Academy of Sciences, Nanjing, 210008, People's Republic of China
        \and
        Astronomy Department, Beijing Normal University, Beijing 100875, China
        \and
        National Astronomical Observatories, Chinese Academy of Sciences\\
   }

   \date{}

\abstract{The Tsinghua University-National Astronomical Observatories of China (NAOC) Transient Survey (TNTS) is an automatic survey for a systematic exploration of optical transients (OTs), conducted with a 60/90 cm  Schmidt telescope at Xinglong station of NAOC. This survey repeatedly covers $\sim$ 1000 square degrees of the north sky with a cadence of 3-4 days. With an exposure of 60 s, the survey reaches a limited unfiltered magnitude of about 19.5 mag. This enables us to discover supernovae at their relatively young stages. In this paper, we describe the overall performance of our survey during the first year and present some preliminary results.
}

   \authorrunning{T. Zhang \& X. Wang}            
   \titlerunning{Tsinghua-NAOC transient survey}  

   \maketitle

%
%
\section{Introduction}           
\label{sect:intro}

Time-Domain Astronomy (TDA) has been recognized as one of the most active and promising research field in astrophysics and is growing rapidly over the past few years. It touches on the understanding of different types of known and unknown transients (or outburst phenomenon) in the universe, such as variables, supernovae, gamma-ray burst, quasars, active galactic nuclei (AGN), and tidal disruption event (TDE) etc.. Wide-field surveys for the transients in the universe open new frontiers in astrophysics.

Owing to diverse scientific objectives, many consortiums have thus put efforts in such surveys by using small- to mediate-size wide-field telescopes, including the Palomar Transient Factory (PTF; \citet{law09}), the La-Silla Quest South Hemisphere Variability Survey (LSQ; \citet{bal13}), the Panoramic Survey Telescope and Rapid Response System (PanSTARRS; \citet{kai02}), the Skymapper Southern Sky Survey \citep{kel07}, and the Catalina Real-Time Transient Survey (CRTS; \citet{drake09}). The next generation large telescopes such as the Large Synoptical Survey Telescope (LSST; \citet{tys03}) will also focus on the time-domain astronomy in the near future.

The Tsinghua University-National Astronomical Observatories of China (NAOC) Transient Survey (TNTS) is an optical survey conducted with a relatively short cadence (e.g., 3-4 days) compared to the existing wide-field transient surveys, aiming primarily at detections of relatively young supernovae in local universe. The early detection enables us to better understand the progenitor system and the explosion physics of SNe. The spectroscopic identifications of our discovery are obtained primarily with Yunan Observatories (YNAO) 2.4-m telescope at Lijiang Station and NAOC 2.16-m telescope at Xinglong Station. The follow-up photometric observations are obtained with the Tsinghua University-NAOC 0.8-m telescope (TNT\footnote{This telescope is co-operated by Tsinghua University and the National Astronomical Observatories of China (NAOC)}) at the Xinglong Station.

In this paper, we present the performance and the first year results of our survey. The observations and data reduction are described in Section 2, and Section 3 presents the results. Our summaries are given in Section 4.


\section{Description of Project}
\label{sect:Description}

\subsection{Instruments}
\label{sect:Description:instruments}

The TNTS is conducted with a Schmidt telescope (with a 90-cm spherical primary mirror and a 60-cm Schmidt corrector plate), located at the Xinglong station of NAOC. One 4096 $\times$ 4096 CCD camera with a plate scale of 1.3 arcsecond per pixel is mounted at the Schmidt focus of the telescope. The field of view (FoV) of the CCD is 90' $\times$ 90'. A detailed description of this telescope was given by \citet{zhou03}. Under a moonless and clear night at Xinglong station, this telescope and the CCD system can reach a detection limit of about 19.5 mag (3 $\sigma$) with the clear filter for an exposure of 60 s. This magnitude limit can detect a normal SN Ia at z $\sim$ 0.04 at about two weeks before its $B$-band maximum light.

\subsection{Survey Strategy}
\label{sect:Description:strategy}

The TNTS is designed to operate for four years starting from October 2012. This survey covers a sky area of $\sim$ 1000 square degrees with Galactic latitude $|b|$ $>$ 10$^{\circ}$ and longitude in the range 0$^{\circ}$ $<$ $\delta$ $<$ 60$^{\circ}$. The nearby galaxy clusters such as the Coma cluster and most part of the Virgo cluster are also included in our survey field, with an intent of catching events of some extremely young supernovae. It usually takes about 2 minutes to take an image for a specific sky field, including 60-s exposure, 22-s readout time, and 30 s for movement and stabilization of the telescope. In order to efficiently rule out the cosmic rays and moving objects, we take two exposures for each sky field with a temporal interval of about 1.0-1.5 hour. The transients with very short timescales of light variations will also benefit from such an observation mode. This means that the whole survey area can be repeatedly visited every 3 to 4 days. Figure 1 shows the sky areas covered by the TNTS, and the red dots are the supernovae candidates discovered during the first-year survey.

\begin{figure}
\centering
\includegraphics[width=70mm, angle=-90]{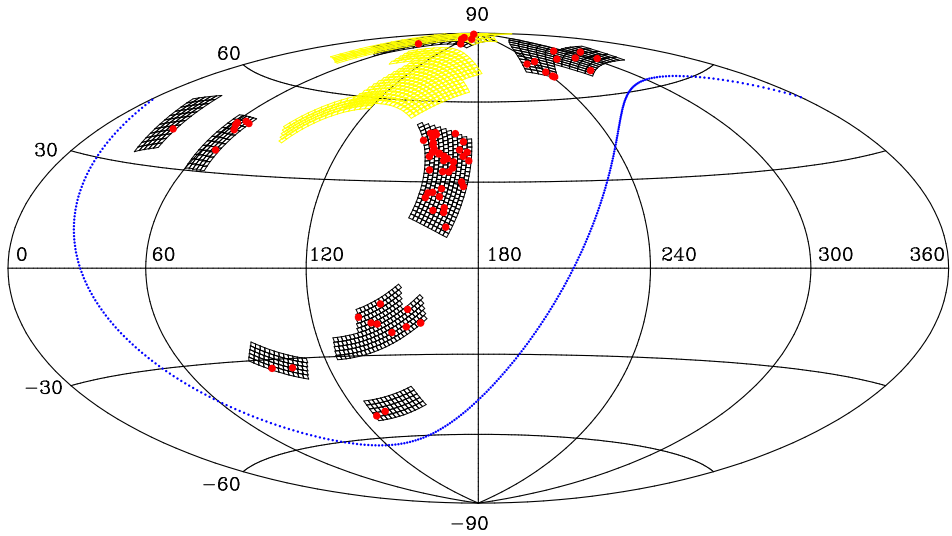}
\caption{The sky area covered by TNTS in Galactic coordinate system. Each small square represents the FoV that the TNTS can cover with one exposure. The yellow regions are the fields that will be covered in the next stage. The red dots represent the SNe candidates discovered by TNTS. The blue line is the celestial equator.}
\label{Fig:1}
\end{figure}

\subsection{Data-Reduction Pipeline}
\label{sect:Description:pipeline}

An image processing pipeline has been developed for the TNTS based on some open source softwares. The software $SExtractor$ is used to find objects and produce the catalogs for each image \citep{bert96}. With these catalogs, the astrometric parameters are obtained by $SCAMP$ \citep{bert06}. The $SWarp$ software is used to resample and align the new images to the reference images \citep{bert02}. After performing the above steps, we apply the image-subtraction technique to detect possible candidates on the residual images. The residual image is obtained by subtracting the reference image from the new image with the High Order Transform of Point Spread Function (PSF) and Template Subtraction ($HOTPANTS$\footnote{See http://www.astro.washington.edu/users/becker/hotpants.html for the details}). This code is effective in detecting the SNe exploding near the central regions of their host galaxies. All these codes and softwares are assembled by $bash$ scripts of linux. Hundreds of sources can be detected on each residual image due to the large field of view of the TNTS, but most can be attributed to artifacts of image subtraction. A series of criteria, such as ellipticity, FWHM, and contamination from bright stars, are applied to rule out the false detections. The remained candidates will be examined carefully by eyes. The most possible candidates will be posted on ATels or CBAT to alert instantly the community to initiate the  follow-up observations.

\subsection{Follow-Up Observations}
\label{sect:Description:followup}

For SNe or other interesting transients discovered by the TNTS at a magnitude brighter than 18.0 mag and are on the rise, we will generally trigger the follow-up observations in photometric and spectroscopic modes.  In particular, extensive follow-up observations will be obtained for those SNe discovered at relatively young phases. The photometric observations are performed with the 0.8-m TNT telescope located at NAOC Xinglong Observatory \citep{wangx08,huang12} in the standard Johnson $UBV$ \citep{john66} and Kron-Cousins $RI$ \citep{cou81} filters, with a 1340 $\times$ 1300 pixel back-illuminated CCD and a FoV of 11.5' $\times$ 11.2' (pixel size $\sim$ 0.52'' pixel$^{-1}$). Spectroscopic observations are taken by the Cassegrain spectrograph and BAO Faint Object Spectrograph \& Camera (BFOSC) mounted on the 2.16-m telescope and Yunnan Astronomical Observatory (YNAO) Faint Object Spectrograph \& Camera (YFOSC) mounted on the 2.4-m telescope at Lijiang Station of YNAO \citep{zhangj12a}. Our goal is to obtain a uniform SN sample with well-sampled $UBVRI$ light curves and spectroscopic data covering premaximum-, maximum-, and post-maximum phases.

\section{The First-Year Results}
\label{sect:Results}

A total of about 30,000 images were obtained during the first-year survey, which yields over 50 SN candidates and lots of other transients such as variable stars, novas, and Quasars/AGNs. 44 SNe finally got spectroscopic identifications from the observations by us and other groups, and the discovery and classifications were published on the CBETs and ATels (See also Table 1).

\begin{figure}
\centering
\includegraphics[width=120mm, angle=0]{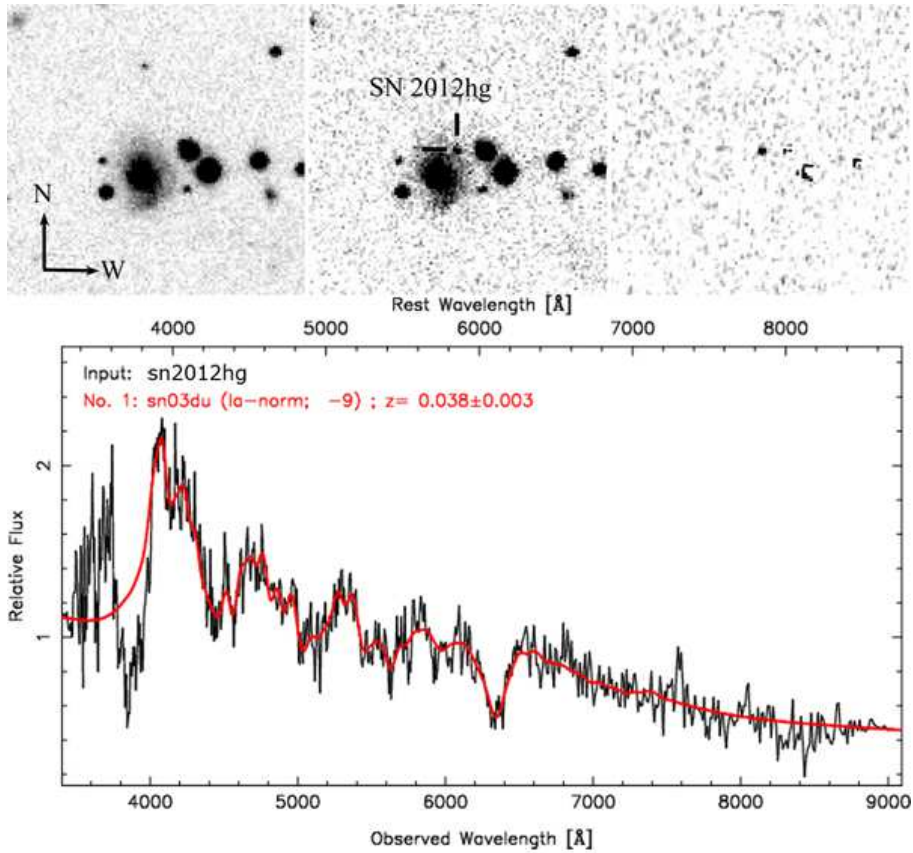}
\caption{The discovery images and spectroscopic identifications of SN 2012hg. Left-up: the template without the SN; Middle-up: the discovery image with the SN; Right-up: the subtracted image with only the SN. Lower pannel: the spectrum of SN 2012hg taken by the 2.4-m telescope. The red line is the best-fit spectrum from SN 2003du at -9 days in the library of SNID \citep{blon07}.}
\label{Fig:2}
\end{figure}

Figure 2 shows the first supernova 2012hg detected by the TNTS. It was discovered on Nov. 25.8 UT at a magnitude of about 18.0 mag. The light curve shows that this SN was actually discovered at about 16 days before its maximum light (see also Figure 5). In Figure 3, we show the spectrum of type IIn SN 2013dw, which is the most distant SN discovered by the TNTS, with a redshift of 0.136 \citep{zhoul13i}. The relevant parameters of the SNe discovered during our first-year survey are listed in Table 1.

\begin{figure}
\centering
\includegraphics[width=120mm, angle=0]{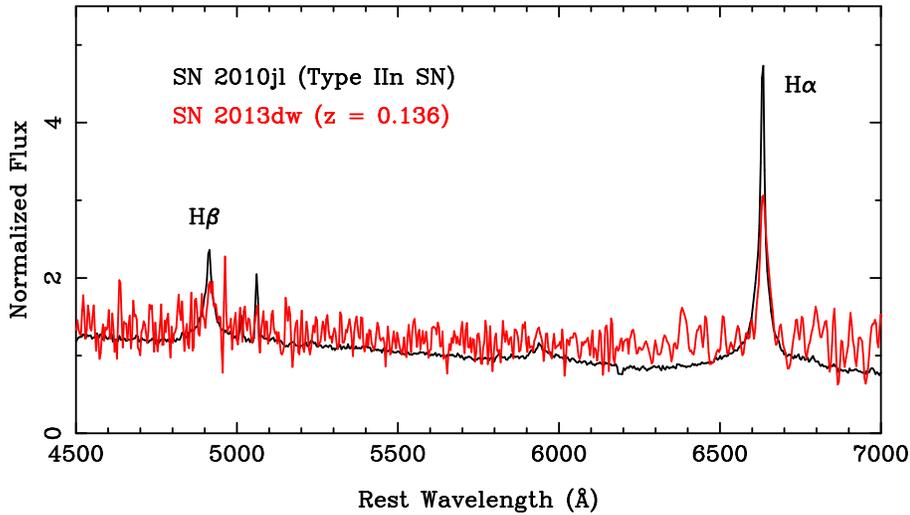}
\caption{The spectrum of bright type IIn SN 2013dw taken by 2.16-m telescope. The redshift of the host galaxy is corrected. The red line is the spectrum of SN 2010jl at about one month after maximum as comparison \citep{zhang12}.}
\label{Fig:3}
\end{figure}

\begin{centering}
\begin{table}
\caption[]{The information of confirmed Supernovae discovered by the TNTS.}\label{Tab:snlist}

{\tiny
\begin{tabular}{l|ccccccc}
  \hline
Designation & Type & R.A. & Dec & Redshift & Discovery Mag & Discovery Date (UT) & Reference \\
  \hline\hline
SN 2012hg & Ia  & 07:07:25.60 & +56:18:19.2 & 0.038 & 18.0 & Nov. 25.8 2012 & CBET 3330~\citep{zhao12}\\
SN 2012hm & Ia  & 02:33:23.32 & +39:40:16.9 & 0.036 & 17.8 & Dec. 07.6 2012	& CBET 3336~\citep{zhangj12b}\\
SN 2012hq & Ia  & 02:07:30.50 & +44:06:19.3 & 0.090 & 18.6 & Dec. 07.7 2012 & CBET 3344~\citep{wangx12}\\
SN 2012hw & IIP & 09:41:38.02 & +48:40:25.5 & 0.038 & 17.9 & Dec. 22.9 2012 & CBET 3353~\citep{howeton12}\\
SN 2012ic & Ia  & 03:13:53.71 & +33:58:03.7 & 0.040 & 17.4 & Dec. 22.7 2012 & CBET 3360~\citep{zhoul12}\\
SN 2012ie & Ia  & 02:24:22.35 & +40:51:03.2 & 0.048 & 18.0 & Dec. 23.5 2012 & CBET 3362~\citep{toma12a}\\
SN 2012ih & Ia  & 06:56:42.91 & +48:54:10.3 & 0.019 & 16.4 & Dec. 10.8 2012 & CBET 3366~\citep{toma12b}\\
SN 2012ii & Ia  & 07:16:55.48 & +51:45:47.6 & 0.060 & 19.0 & Dec. 23.8 2012 & CBET 3369~\citep{zhoul13a}\\
SN 2012ij & Ia  & 11:40:15.84 & +17:27:22.2 & 0.010 & 18.0 & Dec. 31.8 2012 & CBET 3370~\citep{mari13}\\
SN 2012ik & Ia  & 07:35:26.91 & +51:52:50.4 & 0.064 & 19.4 & Dec. 23.8 2012 & CBET 3383~\citep{luppi13}\\
SN 2013N  & Ia  & 11:50:04.13 & +21:16:46.0 & 0.026 & 15.9 & Jan. 26.8 2013 & CBET 3394~\citep{zhoul13b}\\
SN 2013O  & Ia  & 08:52:05.98 & +52:36:06.2 & 0.053 & 18.3 & Jan. 21.7 2013 & CBET 3395~\citep{zhang13a}\\
SN 2013S  & Ia  & 03:35:30.29 & +38:16:59.3 & 0.019 & 16.1 & Jan. 25.6 2013 & CBET 3406~\citep{zhoul13c}\\
SN 2013Z  & IIP & 13:27:54.89 & +30:22:29.4 & 0.050 & 19.0 & Jan. 24.9 2013 & CBET 3415~\citep{inserra13}\\
SN 2013ac & IIP & 09:45:08.79 & +58:40:07.3 & 0.035 & 18.2 & Feb. 15.7 2013 & CBET 3424~\citep{zhang13b}\\
SN 2013af & IIP & 09:13:55.17 & +55:46:56.7 & 0.036 & 18.9 & Mar. 01.5 2013 & CBET 3427~\citep{zhoul13d}\\
SN 2013ah & Ia  & 09:44:33.80 & +55:45:44.4 & 0.025 & 18.6 & Feb. 22.6 2013 & CBET 3430~\citep{elenin13}\\
SN 2013ap & Ia  & 12:58:24.92 & +12:35:53.3 & 0.086 & 19.5 & Feb. 18.9 2013 & CBET 3443~\citep{zhoul13e}\\
SN 2013ar & Ia  & 08:37:45.02 & +49:28:32.2 & 0.052 & 18.9 & Mar. 14.5 2013 & CBET 3446~\citep{zhang13c}\\
SN 2013ax & Ia  & 07:20:03.51 & +55:55:48.4 & 0.040 & 17.4 & Mar. 07.6 2013	& CBET 3455~\citep{zhoul13f}\\
SN 2013be & Ia  & 12:36:27.67 & +11:45:28.1 & 0.066 & 19.6 & Apr. 05.7 2013 & CBET 3470~\citep{silver13}\\
SN 2013bf & Ia  & 08:58:36.07 & +54:19:25.7 & 0.084 & 18.8 & Mar. 28.5 2013 & CBET 3471~\citep{koff13}\\
SN 2013bv & Ic  & 08:41:21.28 & +52:43:30.3 & 0.060 & 18.7 & Apr. 09.5 2013 & CBET 3499~\citep{zhangk13a}\\
SN 2013bx & Ia  & 12:47:24.22 & +32:32:50.0 & 0.078 & 19.8 & Apr. 09.7 2013 & CBET 3501~\citep{zhoul13g}\\
SN 2013ca & IIP & 11:58:43.25 & +19:08:56.2 & 0.043 & 18.1 & May 01.5 2013 & CBET 3508~\citep{zhangk13b}\\
SN 2013cb & Ia  & 11:35:01.74 & +16:07:16.8 & 0.051 & 18.1 & May 01.5 2013 & CBET 3509~\citep{zhang13d}\\
SN 2013co & Ic  & 12:55:50.51 & +30:30:41.5 & 0.050 & 17.7 & May 06.5 2013 & CBET 3527~\citep{zhang13e}\\
SN 2013cp & Ia  & 16:19:52.22 & +38:56:07.9 & 0.075 & 18.5 & May 07.6 2013 & CBET 3528~\citep{zhangk13c}\\
SN 2013cr & Ia  & 16:11:46.47 & +40:51:22.2 & 0.027 & 17.5 & May 14.8 2013 & CBET 3532~\citep{zhang13f}\\
SN 2013cv & Ia  & 16:22:43.16 & +18:57:35.6 & 0.035 & 16.5 & May 20.8 2013 & CBET 3543~\citep{zhoul13h}\\
SN 2013cx & Ia  & 17:04:16.05 & +41:30:37.6 & 0.033 & 17.7 & May 21.7 2013 & CBET 3545~\citep{wangx13}\\
SN 2013dw & IIn & 16:13:58.84 & +42:41:59.0 & 0.136 & 18.8 & Jul. 02.6 2013 & CBET 3585~\citep{zhoul13i}\\
SN 2013ec & Ia  & 16:27:50.26 & +40:28:21.3 & 0.081 & 19.0 & Jul. 02.6 2013 & CBET 3595~\citep{zhangj13a}\\
SN 2013eh & Ia  & 16:16:09.19 & +38:32:53.0 & 0.038 & 16.6 & Jul. 19.5 2013 & CBET 3601~\citep{zhang13g}\\
SN 2013fo & Ia	& 01:20:31.94 & +12:03:12.2 & 0.054 & 17.7 & Sep. 24.6 2013 & CBET 3663~\citep{mo13}\\
SN 2013gf & Ia	& 09:05:26.46 & +56:24:12.4 & 0.100 & 18.3 & Nov. 06.9 2013 & CBET 3702~\citep{zhangj13b}\\
SN 2013gm & IIP & 11:34:21.16 & +15:39:33.1 & 0.017 & 17.5 & Nov. 20.9 2013 & CBET 3726~\citep{zhangj13c}\\
SN 2013gs & Ia  & 09;31:08.87 & +46:23:05.4 & 0.017 & 17.3 & Nov. 29.8 2013 & CBET 3734~\citep{zhangj13d}\\
SN 2013hp & Ia  & 23:13:59.72 & +24:53:38.9 & 0.028 & 18.0 & Dec. 12.4 2013 & CBET 3764~\citep{zhangj13e}\\
\hline
PSN J12541585 & Ia  & 12:54:15.85 & +09:26:25.9 & 0.045 & 17.4 & Feb. 18.3 2013 & ATEL 4808~\citep{cao13}\\
PSN J12393328 & Ic  & 12:39:33.28 & +15:25:52.0 & 0.072 & 19.4 & Feb. 22.8 2013 & ATEL 4851~\citep{wal13}\\
PSN J12533306 & Ia  & 12:53:33.06 & +27:42:51.7 & 0.092 & 19.2 & Mar. 03.8 2013 & ATEL 4860~\citep{nich13}\\
PSN J09040805 & II  & 09:04:08.05 & +47:42:28.0 & 0.047 & 18.8 & Nov. 17.8 2013 & ATEL 5623~\citep{arc13}\\
PSN J12452873 &	II  & 12:45:28.73 & +29:51:04.0 & 0.050 & 18.6 & Dec. 22.8 2013 & ATEL 5700~\citep{cha13}\\
\hline
\hline
\end{tabular}
}

\end{table}
\end{centering}

\begin{figure}
\centering
\includegraphics[width=100mm, angle=0]{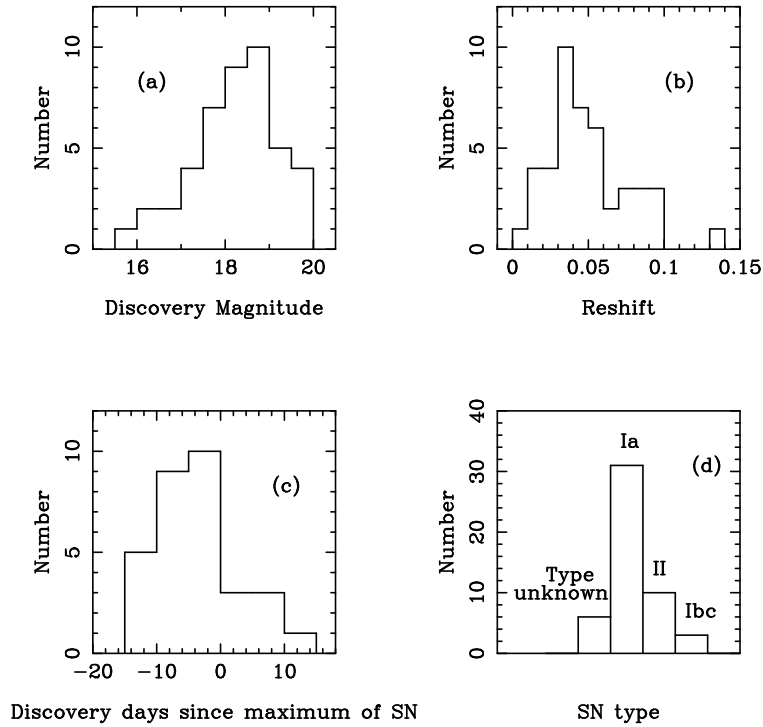}
\caption{The statistic properties of the TNTS SNe. Panel (a): histogram of the SNe magnitude at discovery; Panel (b): the redshift distribution of the TNTS SN sample; Panel (c): histogram of SNe Ia age on their discovery. Panel (d): histogram of SNe type.}
\label{Fig:4}
\end{figure}

\begin{figure}
\centering
\includegraphics[width=80mm, angle=0]{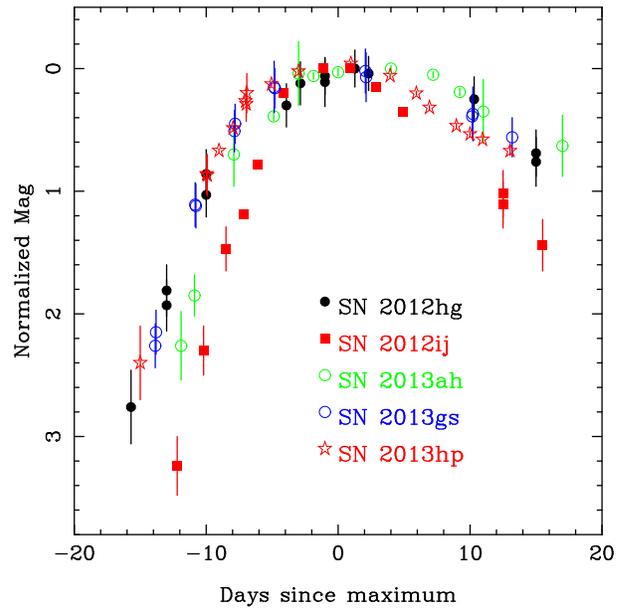}
\caption{The $R$-band (unfiltered) light curves of some young type Ia SNe discovered by the TNTS. All the light curves are normalized to the peak magnitudes and dates for a better comparison.}
\label{Fig:5}
\end{figure}

Figure 4 presents some statistic properties of the first-year SN sample from the TNTS, including the discovery magnitudes, the redshift of the SNe, discovery phase and the spectral types. Figure 4a shows the histogram distribution of the discovery magnitudes of our SN sample. The mean discovery magnitude is $\sim$ 18.2 mag, which is brighter than the detection limit of TNTS by about 1.0 mag. Figure 4b shows the redshift distribution the first-year TNTS SN sample, with a mean value of 0.05. Of this sample, 31 are type Ia SNe, 10 are type II, 3 are type Ibc and 6 are probably SNe without spectral classification (see Figure 4d). The observed fractions of SNe of different spectral types are consistent with those obtained from other magnitude-limited sample (e.g., \citet{lwd11}), which is 17\% for SNe II, 79\% for SNe Ia and 4\% for SNe Ibc, respectively. The higher fraction of SNe Ia is expected in a magnitude-limited survey, as they are on average much brighter than SNe II and SNe Ibc. For SNe Ia, we notice that about 80\% were detected before or around their maximum light, as shown in Figure 4c. The discovery ages of these SNe Ia at discovery were estimated from the unfiltered light curves. Figure 5 illustrates the unfiltered ($R$-band) light curves of five young SNe Ia discovered from the TNTS, which were discovered at about two weeks before the maximum light. These facts prove the capability of TNTS to detect SNe at a relatively young phase. Extensive follow-up observations have also been obtained for these SNe discovered at young phases. The light curves of SN 2012ij indicates that it is a subluminous SN Ia discovered at about 2 weeks before the maximum light, which shows spectroscopic features similar to the subluminous object SN 1991bg \citep{chenj14}. SN 2013gs is another young SN Ia with extensive optical and $Swift$ UV observations. The spectra of SN 2013gs are characterized by high-velocity Si II absorption and it seems to be bright in UV bands \citep{zhangtm14}.

Besides supernovae, a number of other optical transients were also detected during our survey. These include 5 cataclysmic variables (CVs) or novae candidates, 15 active galactic nucleus (AGNs), and about 150 variables \citep{yao14}. Among the 150 variables, 37 are new detections. All of the optical transients discovered by the TNTS can be automatically monitored in an unfiltered mode during the survey. The statistic analysis of variance method, introduced by \citet{stet96}, is used to search for the periodicity of these possible variables. As an example, we show in Figure 6 the phase light curves of two periodic variables found by this method (see \citet{yao14} in details). The preliminary results about the variables prove that, the TNTS has capability to detect variables with different period from hours to months. The survey data can be also used to study the light variations of AGN/quasars. Figure 7 shows the light variations of a flat-spectrum radio quasar (FSRQ) J1310+3233 with a redshift of 1.6 \citep{heal08}. The unfiltered light curve indicates that, the luminosity of this quasar has a long-term rise with a possible variation on a time scale of days. Analysis of this light variation could help to understand the flux contribution of the accretion disk to the quasar emission at different timescales.

\begin{figure}
\centering
\includegraphics[width=100mm, angle=0]{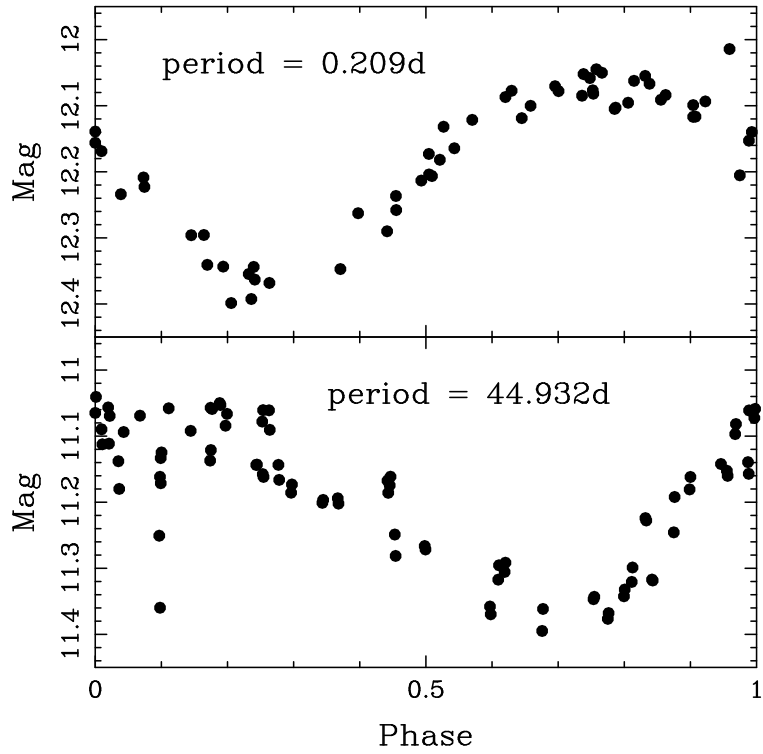}
\caption{Example for two periodic variable stars newly discovered during the survey. Top panel: the unfilter phase light curve with a period of 0.209 days; Bottom panel: phase light curves of another variable with a period of 44.932 days.}
\label{Fig:6}
\end{figure}

\begin{figure}
\centering
\includegraphics[width=100mm, angle=0]{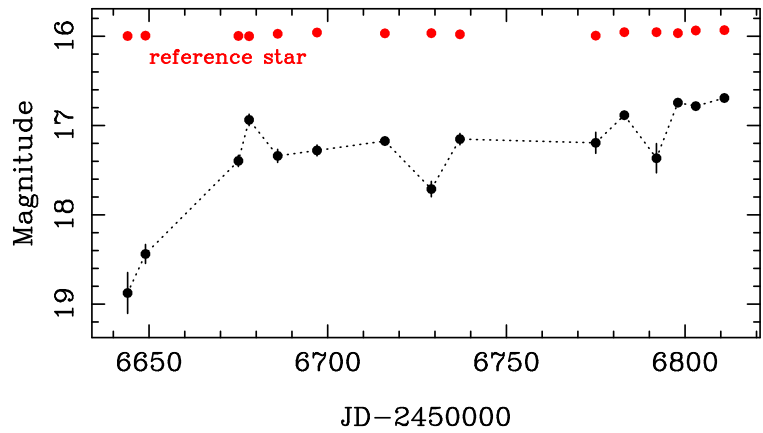}
\caption{The unfiltered light curve of a known quasar FSRQ J131059.4+323334 obtained by the TNTS. The red dots are the light curve of reference star.}
\label{Fig:7}
\end{figure}

\section{Summary}
\label{sect:summary}

This paper introduces the first-year performance of Tsinghua University-NAOC Transient Survey. The observation system and the data reduction pipeline works overall well during the first year survey, and more than 50 SNe and a lot of other transients (e.g., CVs , novas, Quasars/AGNs and variables) have been detected. For some bright SNe and other interesting transients, the photometric and spectroscopic follow-up observations are triggered immediately after their discoveries. From the statistics of the first-year sample of SNe Ia, we found that 80\% of them were discovered before or being close to their maximum light. In particular, it should be pointed that 5 out of 30 SNe Ia were detected at phases around or earlier than two weeks before their maximum light. This number will increase significantly once another two telescopes (one is located at Xinjiang Observatory near Urmiqi and the other is at Xuyi Observatory) join our survey network in the next year. Based on current statistics, our survey can provide the supernova community a sample of above 50 extremely young SNe Ia (e.g., t $<$ -10 days) during its four year operations. Such a sample with early observations will definitely increase our knowledge about SN Ia diversities and their physical origins.

Analysis of a small portion of the survey field leads to the discovery of about 150 variables of different types (including 37 new ones),  with periods ranging from hours to years. From our current data, we estimate that about 1500-2000 variables can be detected from the entire survey fields of the TNTS, which will be a significant contribution to the study of variable stars.

Acknowledgments: This work has been supported by the Chinese National Natural Science Foundation of China (NSFC grants 11203034). The work of X. Wang is supported by NSFC 11178003, 11325313, Tsinghua University Initiative Scientific Research Program, and the Major State Basic Research Development Program (2013CB834903). This work is partially supported by the Strategic Priority Research Program of the Chinese Academy of Sciences, Grant No. XDB09040300, and by the Main Direction Program of Knowledge Innovation of Chinese Academy of Sciences (No. KJCX2-EW-T06), and by the National Basic Research Program of China (973 Program), No. 2013CB834902, 2014CB845704 and 2014CB845702, and by the Chinese National Natural Science Foundation grands No. 11433005, 11073032, 11373035, 11203031, 11303038 and 11303043.

\label{lastpage}

\end{document}